\documentclass[seceq,supplement]{ptptex}

\usepackage{graphicx}
\usepackage{wrapft}

\newcommand{\bv}[1]{{\boldsymbol #1}}
\title{%        %You can use \\ for explicit line-break.
Rheology of sheared granular particles near jamming transition
}

\author{%       %Use \scshape for the family name.
Michio \textsc{Otsuki}$^{1,}$%
and
Hisao \textsc{Hayakawa}$^{2,}$%
}

\inst{%     %Affiliation, neglected when [addenda] or [errata].
$^1$
Department of Physics and Mathematics, Aoyama Gakuin University,
5-10-1 Fuchinobe, Sagamihara, Kanagawa 229-8558, Japan\\
$^2$Yukawa Institute for Theoretical Physics, Kyoto University,  Kitashirakawaoiwake-cho, Sakyo-ku, Kyoto 606-8502, Japan
}

\abst{%         %This abstract is neglected when [addenda] or [errata].
We investigate the rheology of sheared granular materials
near the jamming transition point.
We numerically determine the values of  
the critical fraction and the exponents for the jamming transition using a finite size scaling and 
the nonlinear minimization method known as the Levenberg-Marquardt algorithm.
The exponents are close to our previous theoretical prediction, 
but there is a small discrepancy,
if the critical point is independently determined.
}

\begin{document}

\maketitle

\section{Introduction}

%Jamming

Athermal disordered materials such as colloidal suspensions,\cite{Pusey}
foams,\cite{Durian} and granular materials\cite{Jaeger} 
behave as dense liquids when
the density is lower than a critical value,
while they behave as amorphous solids
when the density exceeds the critical value.
This rigidity transition is known as the jamming transition,
which could be a key concept to characterize disorder materials even for glassy materials.\cite{Liu}

Near the jamming transition point,
such materials show critical behavior,
where the pressure, the elastic moduli,
and the characteristic frequency of the density of state exhibit
power law dependences on the distance 
from the transition point.\cite{OHern02,OHern03,Wyart05}
In particular, the critical scaling law characterized by a set of critical exponents, similar to those in thermal critical phenomena, is observed in the rheology of athermal disordered materials\cite{Hatano07,Olsson07,Olsson11,Tighe,Nordstrom,Hatano08,Hatano10,Otsuki08,Otsuki09,Otsuki10},
though the transition becomes discontinuous under the existence of friction for granular materials.\cite{Otsuki11}.
The precise values of the critical exponents, however,  are still controversial
 because the values of them
are inconsistent among the researchers.\cite{Hatano07,Olsson07,Olsson11,Tighe,Nordstrom,Hatano08,Hatano10,Otsuki08,Otsuki09,Otsuki10,Otsuki11}

In this paper, we try to numerically 
determine the critical exponents near the jamming transition for
 granular materials under the plane shear
 using a nonlinear minimization procedure 
 and a finite size scaling for the critical fraction. 
 The contents of this paper are organized as follows.
Previous results for the critical rheology of athermal disordered materials
are summarized in the next section.
In $\S$ \ref{Estimate:Sec}, the details of our numerical results are presented, where
we explain models and their setup in $\S$ \ref{setup:Sec}, and 
the critical fraction and the exponents are respectively determined in $\S$ \ref{pointJ:Sec} and $\S$ \ref{exponents:Sec}.
In $\S$ 4, we discuss and conclude our results.

\section{Review of scaling properties near the jamming transition}
\label{Rheo:Sec}

Let us consider a sheared athermal system
characterized by the packing fraction $\phi$
and the shear rate $\dot \gamma$.
We restrict our interest to systems consisting of repulsive particles
in which the normal interaction force between contacting particles
is proportional to $\delta_{12}^{\Delta}$
with $\delta_{12} = r_{12} - \sigma_{12}$,
where $r_{12}$ and $\sigma_{12}$ 
is  the distance between the particles'
center of mass and the average diameter of the particles, respectively.
The exponent $\Delta$ characterizes the repulsive interaction, i.e. 
$\Delta = 3/2$ is for spheres of Hertzian contact law,
while the simplified linear model ($\Delta = 1$)
is often used. 
It should be noted that the critical properties are determined by the behavior in the limit of $\delta_{12}\to 0$,
if the repulsive force cannot be characterized by a single $\Delta$.
Thus, 
when the interaction potential analytic near $\delta_{12}=0$, 
such a model always belongs to the same universality class of $\Delta=1$.\cite{Otsuki09}. 
For granular materials,
tangential contact force exists, but is occasionally 
ignored to extract universal properties.
We call the system without the tangential contact force the frictionless system,
while the system with the tangential force the frictional system.

It should be noted that the inertia force is always important for  granular assemblies,
and thus the contact dynamics satisfies an underdamped equation.
On the other hand, the other systems such as foams and colloidal suspensions are believed that
inertia force is negligible and the contact dynamics is described by an overdamped equation.
We also note that granular liquids are characterized by Bagnold's law in which the pressure $P$ and the shear stress $S$
satisfy
\begin{equation}
S  \propto \dot \gamma^2, \quad P  \propto \dot \gamma^2 ,
\label{bagnold}
\end{equation}
while the other liquids such as dense colloids and foams satisfy Newtonian law
\begin{equation}\label{newton}
S  \propto \dot \gamma, \quad P  \propto \dot \gamma .
\end{equation}

In the frictionless athermal systems,
we believe that the jamming transition is continuous.
When the packing fraction  $\phi$ is lower than the jamming fraction $\phi_J$,
which is the onset of the rigidity, the system behaves as a liquid.
Thus, its rheology is characterized by Eq.(\ref{bagnold}) or (\ref{newton}) depending on the system.
When $\phi$ is larger than $\phi_J$,
$S$ and $P$ satisfy
\begin{equation}
S  \propto   (\phi - \phi_J)^{y_\phi}, \quad P \propto (\phi - \phi_J)^{y_\phi '},
\label{yield}
\end{equation}
with the critical exponents $y_\phi$ and $y_\phi '$.
At the critical fraction $\phi_J$,
$S$ and $P$ exhibit power laws as
\begin{equation}\label{power_liquid}
S  \propto   \dot \gamma ^{y_\gamma},\quad P  \propto  \dot \gamma^{y_\gamma '}
\end{equation}
with the critical exponents $y_\gamma$ and $y_\gamma '$.
These rheological properties can be rewritten as the scaling relations
\cite{Olsson07,Hatano08,Otsuki08}
\begin{eqnarray}
S(\dot \gamma, \phi) & = & \dot \gamma^{\beta y_\phi} 
\mathcal{S} \left ( \frac{\phi - \phi_J}{\dot \gamma ^{\beta}} \right ), 
\nonumber \\
P(\dot \gamma, \phi) & = & \dot \gamma^{\beta y_\phi '} 
\mathcal{P} \left ( \frac{\phi - \phi_J}{\dot \gamma ^{\beta}} \right ),
\label{critical_relation}
\end{eqnarray}
with the critical exponent $\beta = y_\gamma / y_\phi =y_\gamma' / y_\phi'$.
Indeed, to satisfy Eqs. (\ref{bagnold}), (\ref{newton}), (\ref{yield}) and (\ref{power_liquid}),
it is sufficient that the scaling functions $\mathcal{S}(x)$ and $\mathcal{P}(x)$ respectively 
satisfy
\begin{eqnarray}
\lim_{x \to \infty} \mathcal{S} (x) \propto x^{y_\phi}, \qquad
\lim_{x \to \infty} \mathcal{P} (x) \propto x^{y_\phi'},
\end{eqnarray}
and
\begin{eqnarray}
\nonumber \\
\lim_{x \to -\infty} \mathcal{S} (x) \propto |x|^{y_\phi - 2 / \beta}, \qquad
\lim_{x \to -\infty} \mathcal{P} (x) \propto |x|^{y_\phi' - 2 / \beta}
\label{scaling_function}
\end{eqnarray}
for the underdamped system,
while
\begin{eqnarray}
\nonumber \\
\lim_{x \to -\infty} \mathcal{S} (x) \propto |x|^{y_\phi - 1 / \beta}, \qquad
\lim_{x \to -\infty} \mathcal{P} (x) \propto |x|^{y_\phi' - 1 / \beta}
\end{eqnarray}
for the overdamped system.
It should be noted that the exponents $y_\phi$ and $y_\phi'$ are believed to be independent of the existence of inertia force.
Indeed, the appearance of the yield stress is determined only by the force transfer in the percolation network of jammed materials.
On the other hand, $y_\gamma$ and $y_\gamma'$ might depend on the detailed properties of dynamics.

Through many simulations and experiments, 
we recognize that there exist some common properties:\cite{Hatano07,Olsson07,Olsson11,Tighe,Nordstrom,Hatano08,Hatano10,Otsuki08,Otsuki09,Otsuki10,Otsuki11}
(i) The critical exponents are insensitive to the spatial dimension if the dimension is above two,
and (ii) the exponents strongly depend on $\Delta$.
These properties are counter intuitive, and is opposite to the conventional critical phenomena.

%Olsson-Teitel

Nevertheless, the values of the critical exponents
are inconsistent among various estimations or observations.
In fact, for overdamped frictionless particles with $\Delta=1$, 
Olsson and Teitel reported
$y_\phi = 1.2$ and $y_\gamma = 0.42$
in their first paper on the jamming transition,\cite{Olsson07}
but in their later paper,\cite{Olsson11}
they estimated $y_\phi = y_\phi' = 1.08$ and $y_\gamma = y_\gamma' = 0.28$.
The theory for overdamped frictionless particles proposed
by Tighe et al.\cite{Tighe} suggests
$y_\phi = \Delta + 1/2$ and $y_\gamma = 1/2$, where they assume that the shear stress 
is given by $S = G \gamma_y$  for $\phi > \phi_J$ with the shear modulus \cite{OHern02,OHern03}
$G \propto (\phi - \phi_J)^{\Delta - 1/2}$
and the yield strain $\gamma_y \propto \phi - \phi_J$, 
which give the prediction of $y_\phi$.
Their prediction is consistent with
the experiment of colloidal suspensions,\cite{Nordstrom}
but contradicts 
with the numerical estimation of Olsson and Teitel.\cite{Olsson07,Olsson11}

%Hatano

For the frictionless granular materials with $\Delta = 1$,
the critical exponents are reported 
as $y_\gamma = 5/7$ and $y_\gamma' = 4/7$ in Ref.~\citen{Hatano07}. 
Hatano found that 
the critical scaling relation \eqref{critical_relation}
holds with $y_\phi= 1.2$, $y_\gamma = 0.63$, 
$y_\phi' = 1.2$, and $y_\gamma' = 0.57$
in his first report,~\cite{Hatano08}
but $y_\phi$ and $y_\gamma$ are respectively estimated
as $1.5$ and $0.6$ in his recent paper.\cite{Hatano10}
Otsuki and Hayakawa proposed a phenomenological theory to predict
$y_\phi = y_\phi' = \Delta$ and 
$y_\gamma = y_\gamma' = 2 \Delta / (\Delta + 4)$,
but the values differ from Hatano's estimation.\cite{Otsuki08}
Note that some of differences between the two groups are superficial.
Indeed, if we use the same $y_\phi$, all exponents in one group agree with those of the other group.
Therefore, the precise estimation of $y_\phi$ is crucial.

For frictional granular systems, which are characterized by a microscopic friction coefficient $\mu$,
the scaling property \eqref{yield} using $\phi_J$
is no longer valid
because the shear stress and the pressure change discontinuously at the jamming point.
However, by introducing a fictitious transition density $\phi_S(\mu)$
depending on the friction coefficient $\mu$,
similar scaling relations exist as
\begin{equation}
S(\phi, \mu) =  A(\mu) \{ \phi - \phi_S(\mu) \}^{y_\phi}
, \quad P(\phi, \mu) = B\{\phi - \phi_S(\mu) \}^{y_\phi'} .
\label{P_QS_scale}
\end{equation}
Otsuki and Hayakawa\cite{Otsuki11} indicate that $y_\phi$ and $y_\phi'$ satisfy
 $y_\phi = y_\phi' = \Delta$, where the prefactor
$A(\mu)$ depends on $\mu$
and $B$ is a constant.\cite{Otsuki11}
We should note that
$\phi_S(\mu)$
coincides $\phi_J$ for the frictionless system.

The estimated values of the exponents 
in the previous papers are summarized in table \ref{Table}.
As shown in the table, the values of the exponents
differ among the papers.
Because we expect that the critical exponents $y_\gamma$ and $y_\gamma '$ characterizing a power law liquid
 depend on the detail of the dynamics,
the differences among $y_\gamma$ and $y_\gamma '$ are quite natural.
We, however, anticipate that the exponents  $y_\phi$ and $y_\phi '$ to characterize the quasi static motion are 
universal.
Thus, the discrepancy among the previous papers on $y_\phi$ and $y_\phi '$ might be a serious problem.

\begin{table}
\caption{The critical exponents reported in the previous papers.
We abbreviate the overdamped system as O, while the underdamped system as U.}
\label{Table}
\begin{center}
\begin{tabular}{cc|cccc} \hline \hline
Paper  & system  & $y_\phi$ & $y_\gamma$  & $y_\phi'$ & $y_\gamma'$ \\ \hline
Olsson and Teitel (2007) \cite{Olsson07} & O (frictionless, $\Delta=1$) & 1.2    &  0.42  & &  \\ 
Olsson and Teitel (2011) \cite{Olsson11} & O (frictionless, $\Delta=1$) & 1.08   &  0.28  & 1.08  & 0.28  \\ 
Tighe, et al. (2010) \cite{Tighe}& O (frictionless) & $\Delta + 1/2$   &  $1/2$ & &  \\ 
Nordstrom, et al. (2010) \cite{Nordstrom}& O (experiment, $\Delta = 3/2$) & 2.1   &  0.48 & &  \\ \hline
Hatano, Otsuki and Sasa (2007) \cite{Hatano07}& U (frictionless, $\Delta=1$) &    &  5/7  & & 4/7  \\
Hatano  (2008) \cite{Hatano08}& U (frictionless, $\Delta=1$)&   1.2 &  0.63  & 1.2 & 0.57 \\
Hatano (2010) \cite{Hatano10}& U (frictionless, $\Delta=1$)&   1.5 &  0.6  &  &  \\ 
Otsuki and Hayakawa (2009) \cite{Otsuki08,Otsuki09}& U (frictionless)& $\Delta$   &  $\frac{2\Delta}{\Delta + 4}$ &  $\Delta$ & $ \frac{2\Delta}  {\Delta + 4}$ \\ 
Otsuki and Hayakawa (2011) \cite{Otsuki11}& U (frictional) 
& $\Delta$   &   &  $\Delta$ &  \\ \hline
\end{tabular}
\end{center}
\end{table}

We should note that the estimation of the exponents depends on the 
choice of the critical fraction $\phi_J$.\cite{Otsuki08,Hatano10}
In Refs.~\citen{Tighe,Otsuki08,Otsuki09}, 
they simultaneously determined $\phi_J$ with the critical exponents.
However, the critical fraction may have to be determined independently as in
Refs.~\citen{Olsson11,Otsuki11}.
In addition, the most of works \cite{Olsson07, Hatano07, Hatano08, Hatano10,Otsuki08,Otsuki09} 
except for Olsson and Teitel\cite{Olsson11} 
did not use
a systematic method,
such as the nonlinear minimization technique known as the Levenberg-Marquardt algorithm\cite{LM}, to estimate the critical exponent.

\section{Numerical result}
\label{Estimate:Sec}

Following Olsson and Teitel,\cite{Olsson11}
we systematically determine the critical exponents 
near the jamming point as well as  the critical fraction $\phi_J$.
In order to determine the critical fraction and the exponents,
we use a nonlinear minimization technique: 
the Levenberg-Marquardt algorithm.

\subsection{Setup}
\label{setup:Sec}

%frictional
Let us consider a two-dimensional granular assembly
in a square box with side length $L$.
%The packing fraction is denoted by $\phi$.
The system includes $N$ grains, each having
an identical mass $m$. The position, velocity, and angular velocity
of a grain $i$ are respectively denoted by
$\bv{r}_i$, $\bv{v}_i$, and $\omega_i$.
Our system consists of grains having 
the diameters $0.7 \sigma_0$, $0.8 \sigma_0$, $0.9 \sigma_0$, 
and $\sigma_0$, where there are $N/4$ for each species of grains.

The contact force $\bv{f}_{ij}$ consists of
the normal part $\bv{f}^{(n)}_{ij}$ and the tangential part $\bv{f}^{(t)}_{ij}$
as $\bv{f}_{ij} = \bv{f}^{(n)}_{ij} + \bv{f}^{(t)}_{ij}$.
The normal contact force $\bv{f}^{(n)}_{ij}$ between the grain $i$
and the grain $j$ is given by
$\bv{f}^{(n)}_{ij} = h^{(n)}_{ij} \Theta(h^{(n)}_{ij})
\Theta (\sigma_{ij} - r_{ij}) \bv{n}_{ij}$,
where $h^{(n)}_{ij}$ and $\bv{n}_{ij}$ are respectively given by
$h^{(n)}_{ij} = k^{(n)} (\sigma_{ij} - r_{ij}) - \eta^{(n)} v^{(n)}_{ij}$
and $\bv{n}_{ij} = \bv{r}_{ij}/|\bv{r}_{ij}|$ with the normal 
elastic constant $k^{(n)}$,
the normal viscous constant $\eta^{(n)}$, 
the diameter $\sigma_i$ of grain $i$,
$\bv{r}_{ij} \equiv \bv{r}_{i} - \bv{r}_{j} $, 
$\sigma_{ij} \equiv (\sigma_i + \sigma_j)/2$ and 
$v^{(n)}_{ij} \equiv (\bv{v}_{i}- \bv{v}_{j}) \cdot \bv{n}_{ij}$. Here, 
$\Theta(x)$ is the Heaviside step function defined by $\Theta(x)=1$
for $x \ge 0$ and $\Theta(x)=0$ otherwise.
Similarly, the tangential contact force
$\bv{f}^{(t)}_{ij}$ between grain $i$
and grain $j$ is given by the equation 
$\bv{f}^{(t)}_{ij} = \min(|h^{(t)}_{ij}|, \mu |\bv{f}^{(n)}_{ij}|) \mathrm{sign} (h^{(t)}_{ij}) \bv{t}_{ij}$,
where $\min(a,b)$ selects the smaller one between $a$ and $b$,
and $h^{(t)}_{ij}$ is given by
$h^{(t)}_{ij} = k^{(t)} u^{(t)}_{ij} - \eta^{(t)} v^{(t)}_{ij}$
with the tangential unit vector 
$\bv{t}_{ij} = (-y_{ij}/|\bv{r}_{ij}|, x_{ij}/|\bv{r}_{ij}|)$.
Here, $k^{(t)}$ and $\eta^{(t)}$ are the elastic and viscous
constants along the tangential direction.
The tangential velocity $v^{(t)}_{ij}$ and the tangential displacement
$u^{(t)}_{ij}$ are respectively given by
$v^{(t)}_{ij} = (\bv{v}_{i}- \bv{v}_{j}) \cdot \bv{t}_{ij}
+ (\sigma_i \omega_i + \sigma_j \omega_j)/2$
and $u^{(t)}_{ij} = \int_{\mathrm{stick}} dt \  v^{(t)}_{ij}$,
where  ``stick'' on the integral indicates that
the integral is performed 
when the condition $|h^{(t)}_{ij}| < \mu |\bv{f}^{(n)}_{ij}|$ or 
 another condition $u^{(t)}_{ij} v^{(t)}_{ij} < 0$
is satisfied.\cite{DEM, Hatano09}

We investigate the shear stress $S$ and the pressure $P$,
which are respectively given by 
\begin{eqnarray}
S & = &  -\frac{1}{L^2}\left <    \sum_i^N \sum_{j>i} 
r_{ij,x}
 \left [ f^{(n)}_{ij,y} + f^{(t)}_{ij,y} 
 \right ]
\right > 
%\nonumber \\
%& &
%-\frac{1}{V} \left <   \sum_{i=1}^N \frac{p_{x,i}p_{y,i}}{2m} \right >
\label{S:calc},  \\
P & = & 
\frac{1}{2L^2} \left < \sum_i^N \sum_{j>i} \bv{r}_{ij} \cdot
 \left [ \bv{f}^{(n)}_{ij} + \bv{f}^{(t)}_{ij} \right ]
\right > ,
%& & +
%\frac{1}{2V} \left <   \sum_{i=1}^N \frac{|\bv{p}_i|^2}{2m} \right >,
\label{P:ex}
\end{eqnarray}
where 
$\left < \cdot \right >$ represents the ensemble average.
Here, we ignore the kinetic parts of $S$ and $P$,
which are respectively given by
$S_{\rm K} =  - \left <    \sum_i^N
p_{i,x} p_{i,y} \right > / (mV)$ and $P_{\rm K} = 
\left <    \sum_i^N
\bv{p}_{i} \cdot \bv{p}_{i}  \right >/ (2mV)$,
because they are significantly smaller than the potential parts in Eqs. (\ref{S:calc}) and (\ref{P:ex})
near the jamming transition point.

In this paper, the shear is imposed 
along the  $y$ direction and macroscopic displacement 
only along the $x$ direction by the following two methods.
The first method is the SLLOD algorithm
under the Lees-Edwards boundary condition \cite{Evans}, which we call ``SL'' for later discussion,
where the time evolution is determined by
\begin{eqnarray}
\frac{d \bv{r}_i}{dt} & = & \frac{\bv{p}_i}{m} + \dot \gamma y_i \bv{e}_x,
\label{SLLOD:1} \\
\frac{d \bv{p}_i}{dt} & = & \sum_{j \neq i} \bv{f}_{ij} - 
\dot \gamma p_{i,y} \bv{e}_x
\label{SLLOD:2}
\end{eqnarray}
with the peculiar momentum $\bv{p}_i = m (\bv{v}_i - \dot \gamma y \bv{e}_x)$
and the unit vector parallel to the $x$-direction $\bv{e}_x$.

The second method is {\it quasi-static shearing method}, which we call ``QS''. 
\cite{Heussinger09,Vagberg10}
In this method,
the shear strain $\Delta \gamma$ is applied by an affine transformation
of the position of the particles.
Then, the particles are relaxed under the time evolution equations
\begin{eqnarray}
\frac{d \bv{r}_i}{dt} & = & \frac{\bv{v}_i}{m}, \\
\frac{d \bv{v}_i}{dt} & = & \sum_{j \neq i} \bv{f}_{ij}
\end{eqnarray}
until the kinetic energy per particles becomes lower than
a threshold value $E_{\rm th}$.
Then, we repeat applying the shear and the relaxation process.
Here, we chose $\Delta \gamma = 10^{-6}$ and $E_{\rm th} = 10^{-7} k^{(n)}\sigma_0^2$,
which are small enough not to influence our results.
This method is expected to correspond to the low shear limit of the SL method.

In our simulation $m$, $\sigma_0$ and $k^{(n)}$ are set to be unity, 
and all quantities are converted to dimensionless forms,
where the unit of time scale is $\sqrt{m / k^{(n)}}$.
We use the viscous constants $\eta^{(n)} = \eta^{(t)} = 1.0$
and the tangential spring constant $k^{(t)} = k^{(n)}$ for the frictional case.

\subsection{Determination of the critical fraction for the frictionless case}
\label{pointJ:Sec}

\begin{figure}
\begin{center}
\includegraphics[width=0.5\textwidth]{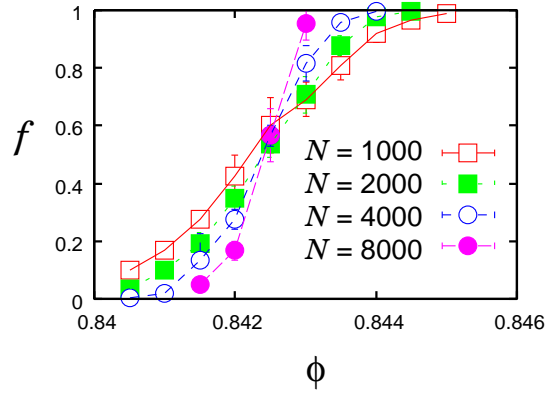}
\caption{
(Color online)
Jammed fraction $f$ as a function of $\phi$ for $N=1000, 2000, 4000$ and $8000$.
}
\label{f_fig}
\end{center}
\end{figure}

\begin{figure}
\begin{center}
\includegraphics[width=0.5\textwidth]{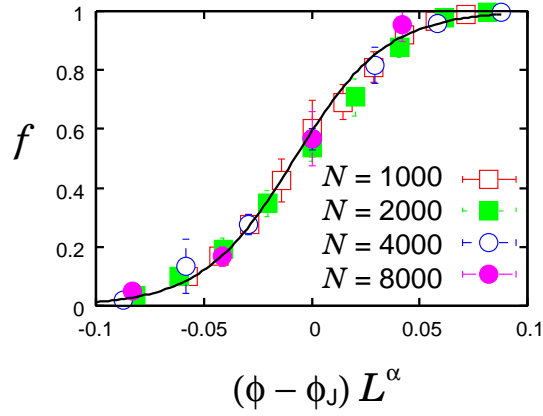}
\caption{
(Color online)
  Scaling plots of the 
jammed fraction $f$ characterized by Eq. \eqref{f_scale}.
The solid line is the scaling function given by Eq. \eqref{f_sfunc}.
}
\label{f_scale_fig}
\end{center}
\end{figure}

In this subsection, we determine the transition density $\phi_J$ for frictionless particles
by introducing the jammed fraction $f$ obtained from the simulation
using the QS method.
Here, $f$ is the fraction of samples where
the shear stress $S$ is larger than a threshold value $S_{\rm th} = 10^{-6}$.
Figure \ref{f_fig} demonstrates the jammed fraction $f$ as a function of $\phi$.
$f$ is zero in the low density region, while $f$
has a finite value when $\phi$ is large enough,
which suggests the appearance of the yield stress and the rigidity.
It is to be noted that $f$ around $\phi=0.8425$ becomes steeper
as the system size increases.
In order to determine $\phi_J$ from the data in Fig. \ref{f_fig},
we assume $f(\phi, L)$ satisfies
a scaling relation
\begin{equation}
f(\phi, L) = F\left ( (\phi - \phi_J){L^{\alpha}} \right )
\label{f_scale}
\end{equation}
with an exponent $\alpha$ and a scaling function $F(x)$,
which satisfies
$\lim_{x \to \infty} F(x) = 1$
and 
$\lim_{x \to -\infty} F(x) = 0$.
Figure \ref{f_scale_fig} shows the scaling plot based on Eq. \eqref{f_scale}.
This figure confirms the validity of the scaling relation \eqref{f_scale}.
Here, we numerically estimate $\phi_J = 0.84250 \pm 0.00004$, 
where we assume
the functional form of the scaling function as
\begin{equation}
F(x) = \left \{ 1 + \tanh \left ( \frac{x+b}{\Delta x}\right ) \right \} / 2
\label{f_sfunc}
\end{equation}
with the fitting parameters 
$b = 0.0079, \Delta x = 0.042$, and $\alpha = 1.0$.
Note that the critical fraction $\phi_J=0.84250\pm 0.00004$ is almost identical to the 
simultaneously determined value $0.84260\pm 0.0004$ with the critical exponents~\cite{Otsuki08,Otsuki11}.
However, as will be shown, this slight difference between two critical fractions affects the value of the critical exponents.

\subsection{Determination of the critical exponents}
\label{exponents:Sec}

\begin{figure}
\begin{center}
\includegraphics[width=0.9\textwidth]{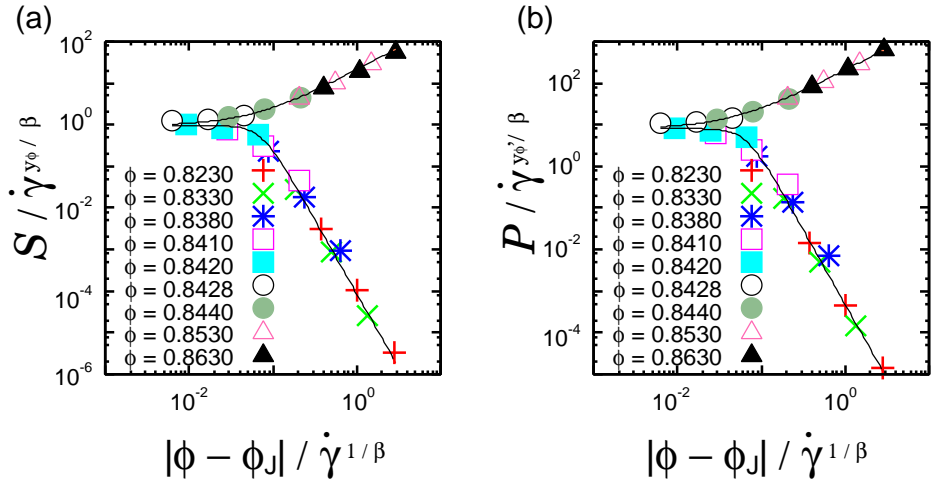}
\caption{
(Color online) 
(a): Scaling plot of the 
shear stress $S(\dot \gamma, \phi)$ for the frictionless systems
characterized by Eq. \eqref{critical_relation}.
The solid line is the scaling function given by Eq. \eqref{sfunc:S}.
(b): Scaling plot of the 
pressure $P(\dot \gamma, \phi)$ for the frictionless systems
characterized by Eq. \eqref{critical_relation}.
The solid line is the scaling function given by Eq. \eqref{sfunc:P}.
Both plots are obtained for case A in which the critical fraction is independently determined.
}
\label{frictionless}
\end{center}
\end{figure}

In this subsection, let us determine the critical exponents
from the simulation of the sheared frictionless system
using the SL method for $5.0 \times 10^{-7} \le \dot \gamma \le
5.0 \times 10^{-5}$ with $N=4000$.
Here, we have determined the critical exponents independently, in which the critical fraction has been determined as in the previous subsection (case A).
Figure \ref{frictionless}
shows the scaling plots 
of $S$ and $P$ based on Eq. \eqref{critical_relation}.
This figure confirms the validity of the scaling relation 
\eqref{critical_relation}.
Here, we numerically determine 
$y_\phi = 1.09 \pm 0.04$,
$y_\phi' = 1.06 \pm 0.04$,
and
$\beta = 0.43 \pm 0.01$,
where we assume the functional forms of the scaling functions as
\begin{eqnarray}
{\mathsf S} (x) & = & S_0 (1+ A_s x^{y_\phi}) \theta(x) 
 + S_0 /(1+ B_s |x|^{ 2 / \beta - y_\phi}) \theta(-x), 
\label{sfunc:S}\\
{\mathsf P} (x) & = & P_0 (1+ A_p x^{y_\phi'}) \theta(x) 
 + P_0 /(1+ B_p |x|^{ 2 / \beta - y_\phi'}) \theta(-x), 
\label{sfunc:P}
\end{eqnarray}
which satisfy Eq. \eqref{scaling_function}
with fitting parameters
$S_0              = 0.96$,
$P_0              = 8.0$,
$A_s              = 21$
$A_p              = 24$,
$B_s              = 11157$,
and
$B_p              = 16803$.
The estimated values are close to the prediction,
$y_\phi = 1.0$, $y_\phi ' = 1.0$, and $\beta = 0.4$
by Otsuki and Hayakawa,\cite{Otsuki08}
but a small discrepancy exists.
It should be noted that Olsson and Teitel reported
the exponents $y_\phi = y_\phi ' = 1.08$ in Ref.~\citen{Olsson11}, 
which is close to our results.

On the other hand, we evaluate the critical exponents $y_\phi=1.0\pm 0.1$, $y_\phi'=1.0\pm 0.1$, $\beta=0.40\pm 0.01$, $y_\gamma=0.40$ and $y_\gamma'=0.40$ if we simultaneously determine both the critical exponents and the critical fraction (case B).
We should stress that the exponents for case B are identical to those obtained from the mean field theory.\cite{Otsuki08}
It is remarkable that the difference of the critical fraction which is about 0.01 $\%$ affects the values of the critical exponents.
We believe that the exponents for case A are more appropriate for the critical scaling than those for case B, 
because the critical exponents are only defined in the vicinity of the true critical fraction which can be determined independently.  
In table \ref{Table2}, we compare the exponents for case A and case B.

\begin{figure}
\begin{center}
\includegraphics[width=0.9\textwidth]{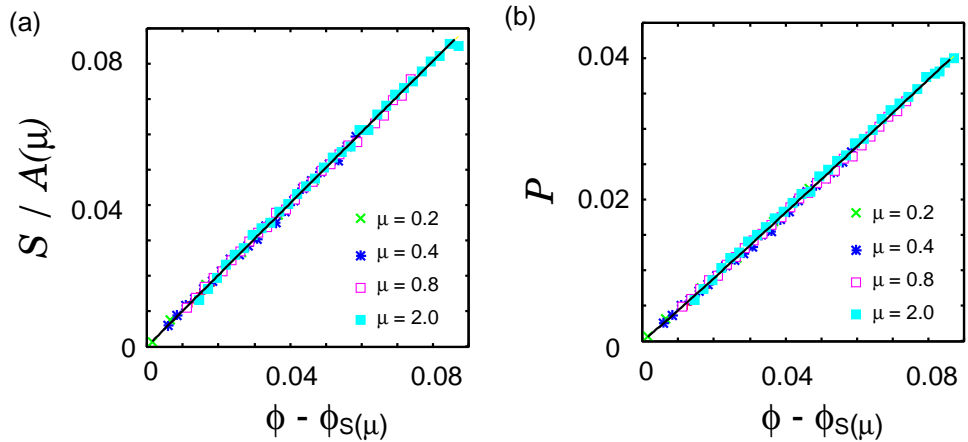}
\caption{
(Color online)
(a) :  Scaling plot of the 
shear stress $S(\dot \gamma, \phi)$
characterized by Eq. \eqref{P_QS_scale}.
(b) :  Scaling plot of the 
pressure $P(\dot \gamma, \phi)$
characterized by Eq. \eqref{P_QS_scale}.
}
\label{frictional}
\end{center}
\end{figure}

For the frictional systems, we can only use case B to determine the critical exponents
based on Eq. \eqref{P_QS_scale}
from the simulation using the SL method
with the shear rate $\dot \gamma = 5.0 \times 10^{-6}$
and $N=4000$.
This is because the jamming transition for frictional grains is discontinuous and the critical exponents are only fictitious ones. 
The estimated values are
$ y_\phi             = 0.97          \pm 0.01$ and 
$y_\phi '              = 0.98          \pm 0.01$
with the fitting parameters
$\phi_S(\mu = 0.2)          = 0.82$,
$\phi_S(\mu = 0.4)         = 0.81$,
$\phi_S(\mu = 0.8)         = 0.79$,
$\phi_S(\mu = 2.0)         = 0.78$,
$A(\mu = 0.2)            = 0.10$,
$A(\mu = 0.4)            = 0.11$,
$A(\mu = 0.8)            = 0.12$,
$A(\mu = 2.0)            = 0.12$,
and 
$B           = 0.44$.
The estimated exponents are almost identical to those in the previous prediction\cite{Otsuki11} and those of frictionless grains for case B.
%The reason for the discrepancy between the frictional case and the frictionless case will be explained in the next paragraph.
Figure \ref{frictional} shows the scaling plots 
of $S$ and $P$ for the frictional particles
based on Eq. \eqref{P_QS_scale},
which verifies the validity of the estimation.

%It is notable that we can simultaneously determine both the critical exponents and the critical fraction 
%from the data of the sheared granular systems.
%This method is used for the frictional case, because we should use the fictitious critical point $\phi_S(\mu)$ instead of the true jamming point.
%In table \ref{Table2}, we compare the exponents
%determined with the critical fraction obtained in $\S$ \ref{pointJ:Sec},
%which we call  case A,
%to those determined with $\phi_J$ as a fitting parameter,
%which we call  case B.
%It is surprised that the exponents for frictionless systems of case B are almost identical to those in both the frictional case and the prediction by Otsuki and Hayakawa\cite{Otsuki08}, though the values
%slightly deviate from those for case A.

\begin{table}
\caption{The critical exponents determined by using a nonlinear minimization
method. 
(Case A) : The exponents are determined with $\phi_J$ obtained in $\S$ 
\ref{pointJ:Sec}.
(Case B) : The exponents are simultaneously determined with $\phi_J$.
}
\label{Table2}
\begin{center}
\begin{tabular}{c|ccc|cc|c} \hline \hline
 & $y_\phi$    &  $y_\phi'$ & $\beta$ &$y_\gamma$ &$y_\gamma$' & $\phi_J$ \\ 
 \hline
Frictionless (case A) & $1.09 \pm 0.04$   &  $1.06 \pm 0.04$ &$0.43 \pm 0.01$  &  $0.47$ & $0.46$ & $0.84250 \pm 0.00004$ \\ 
Frictionless (case B) & $1.0 \pm 0.1$   &  $1.0 \pm 0.1$ &$0.40 \pm 0.01$  &  $0.40$ & $0.40$ & $0.84260 \pm 0.0004$ \\ 
\hline
Frictional (case B) & $0.97 \pm 0.01$   &  $0.98 \pm 0.01$ &  &   &  & \\ 
 \hline
\end{tabular}
\end{center}
\end{table}

\section{Discussion and conclusion}
\label{Discussion:Sec}

Let us compare our results with those of the previous papers.
Tighe et  al. predicted $y_\phi = 1.5$ for the system with $\Delta = 1$,
which is consistent with the numerical results for 
overdamped \cite{Tighe}
and underdamped systems.\cite{Hatano08,Hatano10}
However, they did use any systematic method, such as the nonlinear minimization technique for the determination of the critical exponents.
Our systematic determination of the critical exponents
for the frictionless granular system gives e.g.
$y_\phi = 1.09 \pm 0.04$ for cae A while case B where the critical exponents are simultanaously determined with the critical fraction gives $y_\phi=1.0\pm 0.1$.
We should note that $y_\phi$ for case A
is almost identical to another systematic estimation for an overdamped system.\cite{Olsson11}
It still remains possibility that  the estimation depends on the range of the shear rate
and the density.~\cite{Hatano10,Otsuki09}
However, our new result for case A may support the suggestion\cite{Olsson11} that $y_\phi$ is close but slightly larger than 1.
We also note that the previous exponents in terms of the mean field theory are almost identical those for case B.
It is likely that the deviation from the mean field prediction is significant to represent the existence of critical fluctuations.

We should note that the critical scaling of the jamming transition for frictional grains is fictitious, because the actual transition is discontinuous. 
For frictional systems, thus, we can only use case B, in which the exponents are almost identical to those for the frictionless case.

In conclusion, we numerically determined the critical exponents
for the jamming tranistion of granular materials near the jamming transition point.
The estimated values for case A are close to the previous theoretical prediction\cite{Otsuki08} and those for case B
but a small deviation exists
for the frictionless system.
The value of case A is almost identical to those obtained for the rheology
of foams near the transition point.\cite{Olsson11}
The fictitious critical exponents for frictional grains are almost identical to those for case B and the theoretical prediction of the frictionless grains.

\section*{Acknowledgments}
We thank S. Teitel for valuable discussions.
This work is partially supported by the 
Ministry of Education, Culture, Science and Technology (MEXT), Japan
 (Grant Nos. 21540384 and 22740260) and the Grant-in-Aid for the global COE program
"The Next Generation of Physics, Spun from Universality and Emergence"
from MEXT, Japan.
The numerical calculations were carried out on Altix3700 BX2 at 
the Yukawa Institute for Theoretical Physics (YITP), Kyoto University.

%\appendix
%\section{First Appendix} %Empty argument \section{} yields `Appendix'. 
%
%\section{Second Appendix}

\end{document}